# Building a Decision Support System for Automated Mobile Asthma Monitoring in Remote Areas

Uwaoma, Chinazunwa, The University of the West Indies, Jamaica
chinazunwa.uwaoma@mymona.uwi.edu

Mansingh, Gunjan, The University of West Indies, Jamaica
gunjan.mansingh@uwimona.edu.jm

**ABSTRACT**

Advances in mobile computing have paved the way for the development of several health applications using Smartphone as a platform for data acquisition, analysis and presentation. Such areas where m-health systems have been extensively deployed include monitoring of long-term health conditions like Cardio-Vascular Diseases and pulmonary disorders, as well as detection of changes from baseline measurements of such conditions. Asthma is one of the respiratory conditions with growing concern across the globe due to the economic, social and emotional burden associated with the ailment. The management and control of asthma can be improved by consistent monitoring of the condition in real-time since attack could occur anytime and anywhere. This paper proposes the use of smartphone equipped with built-in sensors, to capture and analyse early symptoms of asthma triggered by exercise. The system design is based on Decision Support System (DSS) techniques for measuring and analysing the level and type of patient's physical activity as well as weather conditions that predispose asthma attack. Preliminary results show that smartphones can be used to monitor and detect asthma symptoms without other networked devices. This would enhance the usability of the health system while ensuring user's data privacy, and reducing the overall cost of system deployment. Further, the proposed system can serve as a handy tool for a quick medical response for asthmatics in low-income countries where there are limited access to specialized medical devices and shortages of health professionals. Development of such monitoring systems signals a positive response to lessen the global burden of asthma.

**Keywords:**

 Decision support, Smartphone, Asthma monitoring, Global burden, M-health, Development

1. **INTRODUCTION**

Recent years have witnessed a paradigm shift from the conventional use of cellular phones in supporting only voice communication to multimedia services and healthcare applications. The massive availability and improvement on the storage, networking and computing capabilities of these devices have allowed the development of a wide range of mobile applications including healthcare systems. Combination of mobile computing technologies and wireless body sensor networks (WBSN) presents opportunities for unobtrusive monitoring of patient's clinical data and seamless communication of the monitored data to healthcare professionals. Asthma is one of the long-term health conditions whose treatment can benefit from real-time and continuous monitoring, as an attack could occur anytime and anywhere.

Braman (2006) reports that poor asthma control is significantly responsible for low productivity among workers, job losses and absenteeism from school in both developed and developing countries. According to WHO statistics, over 3000 million people worldwide suffer from Asthma while more than 250, 000 die each year from asthma cases (Braman, 2006; Masoli et al., 2004). The rising increase in asthma cases has necessitated the development of technological support for asthma management and control (Seto et al., 2009; Uwaoma & Mansingh, 2014). Collaborative efforts have also been made towards reducing the global burden of Asthma. About a decade ago, the Global Initiative for Asthma (GINA) came up with guidelines for global asthma management and control which include objective assessment of the condition, patient-doctor partnership, control of environmental influences, patients maintaining normal activities/exercises, and adherence to medications (Braman, 2006; Masoli et al., 2004). However, the goals of this Initiative cannot be fully achieved without considering certain barriers observable in developing countries such as insufficient access to medical care and essential drugs, and inadequate communication between patients and their care providers.

Asthma manifests in variant forms which can be categorized according to the trigger factors and temporal-spatial dynamics of the attack. Our study focuses on Exercise-induced Asthma (EIA) - an asthma flare due to strenuous physical exertion. McFadden and Gilbert (1994) observe that asthmatics with chronic conditions manifest signs of asthma attack during exercise. Also, there are many people without asthma who develop symptoms of bronchospasm only during rigorous exercises like sporting and prolonged strenuous activities common among rural dwellers. However, medical experts have advised that asthma should not be an excuse to deprive persons the benefits of physical exercise since asthmatics can lead normal and productive life if the condition is properly handled (Milgrom & Tausssig, 1999). But asthma patients do not always report early signs of asthma flare which may result in further exacerbation of the condition.

EIA is associated with wheeze and shortness of breath. Person experiencing asthma attack also tends to lean forward in an effort to get sufficient air into the lungs which invariably makes the patient to assume an inclined posture (Signs of a pending Asthma Attack n.d). Wheezing is the most investigated vital sign among the common symptoms of asthma as it is reported to be a major predictor to asthma condition (Braman, 2006). However, not all wheezes points to asthma and not all asthmatics presents wheeze during an episode; hence, the need to include contextual information to help evaluate patient's condition. Furthermore, the identification of ambient conditions associated with pathophysiology of asthma as well as measuring patient's level of physical activities (Karantonis et al., 2006) on detection of any anomaly in breath pattern, could help in real-time medical intervention for asthmatics. This paper proposes a design for asthma monitoring based on embedded intelligent systems in smartphones, to assist asthma patients and their care givers in managing asthma flares induced by rigorous activities.

The rest of the paper is structured as follows. Section 2 provides background and enabling concepts for our design. In section 3, we discuss the methodology and present an overview of the design architecture in section 4. Design implementation and results from the preliminary testing are discussed in section 5 while application of the proposed system is highlighted in section 6. Section 7 summarises our discussion and the direction for future work.

## 2. BACKGROUND

There has been notable increase in the use of advance information and communications technologies to improve the overall asthma management and control. This spans from electronic peak-flowmetry and asthma diaries through asthma web-based tools to mobile phone applications (Glykas & Chytas, 2004; Hendler et al., 2012; Uwaoma & Mansingh, 2014). Building on their previous work – DexterNet, Seto et al., (2009) proposed an architecture for comprehensive asthma monitoring. The system consists of a mobile device as a sink for the aggregation and processing of clinical and physiological data. It extends the functionalities of the existing asthma e-health systems to accommodate monitoring of physical activity and outdoor exposures to environmental triggers. The architecture also includes web-based applications which provides platform for collaborative work among the patients, healthcare providers and health researchers.

Lung sound is one of the most vital signs monitored in asthma given that its analysis provides medical doctors with critical information on how to adjust treatment for patients with asthma condition. Thus, detection and analysis of abnormal respiratory sounds like wheezes becomes paramount in the design of asthma e-health systems. Wisniewski and Zielinski (2010) argue that asthma e-health designs are not sufficient to provide comprehensive monitoring of asthma patients without the inclusion of lung sound analysis. The authors advance a 'fully' integrated asthma e-health system that will not only allow for detection of asthma wheeze but also include measurement of other vital signs and triggering factors.

Asthma wheeze detection systems are developed alongside classification models which make it possible to precisely qualify and distinguish adventitious sound from normal respiratory sounds (Shaharum et al. 2012; Reichert et al. 2008; Taplidou & Hadjileontiadis, 2007, Uwaoma & Mansingh, 2014). Several approaches adopt machine learning techniques using advanced classifiers rather than threshold-based algorithms, to produce more elaborate and reliable results. Oletic et al. (2014), evaluated different wheeze detection algorithms based on Decision Tree classification. The aim is to determine which metrics has the best classification accuracy and efficiency that could be implemented on a low-power wearable device for automated recognition of respiratory sound patterns.

Knowledge-based and Decision Support Systems for mobile healthcare are built to detect and alert on any anomaly, and also provide summary report on monitored health condition. These intelligent systems use context information from different sources to correlate data from a monitored event in order to arrive at a decision (Minutolo et al., 2010). In situation or context aware systems, extracted data from different sources are often validated using a set of conditional tests. Usually, a chain of conditions are applied to the given data to identify the condition(s) that matches the required pattern. These set of conditional tests are called production rules and can be combined with ontologies for efficient processing of information. An ontology provides formal framework for organization and representation of domain knowledge (Henriques et al., 2013).

In our study, we leverage the potentials of embedded sensors in smartphones for data collection, and the phone itself providing the platform for data processing, analysis, presentation and feedback (Uwaoma & Mansingh, 2015). This is enhanced by incorporating contextual information through machine learning techniques and expert systems to assist in decision process. The ability of the system to automatically signal warnings on detection of abnormalities and to generate feedback would help to improve patient's self-management and communication with health professionals. Also, by utilizing the capabilities of internal sensors, modern mobile phones can serve as veritable assistive tools for monitoring and alerting asthma patients and their care providers on early symptoms of asthma exacerbation. This removes the need of having external sensors and other networked devices attached on the user, which could increase installation cost and compromise user's data which by default are sent over to a remote server for computation and analysis. In addition, since the proposed system would be deployed as a standalone monitoring device, the unavailability of internet connectivity and mobile data services in remote and rural areas will not pose so much a challenge to the overall performance of the system.

## 3. METHODOLOGY

In this section we briefly highlight the general principles that guide our proposed system, and then discuss the methods we adopt to realize the system.

Studies that center around information systems can be approached in two perspectives namely; research as a *behavioral science* and research as a *design science* (Hevner, 2004). Behavioral science addresses research through the development and justification of theories that explain or predict phenomena related to the identified organizational need. Design science on the other hand, involves building and evaluation of artifacts designed to meet the identified need. Both approaches are distinct but complement each other. While the goal of behavioral science research is *truth*, the goal of design science research is *utility* - truth and utility are inseparable. Truth informs design and utility informs theory (Hevner, 2004).

Though generally, researchers will prefer one approach to the other, at times, there are situations that call for a combination of both depending on the research phenomena being investigated. In our study the phenomena being considered are two-pronged: one being 'man-made' (the mobile device), and the other (asthma health condition) having a relatively natural occurrence. The scenario here essentially warrants a hybrid of methods from both behavioral science and design science to justify and evaluate the proposed system. However, our approach tilts more towards design science which is fundamentally, a problem-solving paradigm focusing on "the creation of new and innovative artifacts" to improve the existing practices in the problem domain (Hevner,

2004). An artifact can be in form of a construct, model, method or instantiation. Our study is not only proposing a model but we are also instantiating the model in order to evaluate its applicability in real-world situations.

| Guideline | Description |
| --- | --- |
| Guideline 1: Design as an Artifact | Design-science research must produce a viable artifact in the form of a construct, a model, a method, or an instantiation. |
| Guideline 2: Problem Relevance | The objective of design-science research is to develop technology-based solutions to important and relevant business problems. |
| Guideline 3: Design Evaluation | The utility, quality, and efficacy of a design artifact must be rigorously demonstrated via well-executed evaluation methods. |
| Guideline 4: Research Contributions | Effective design-science research must provide clear and verifiable contributions in the areas of the design artifact, design foundations, and/or design methodologies. |
| Guideline 5: Research Rigor | Design-science research relies upon the application of rigorous methods in both the construction and evaluation of the design artifact. |
| Guideline 6: Design as a Search Process | The search for an effective artifact requires utilizing available means to reach desired ends while satisfying laws in the problem environment. |
| Guideline 7: Communication of Research | Design-science research must be presented effectively both to technology-oriented as well as management-oriented audiences. |

*Table 1. Design Science Research Guidelines (Hevner, 2004)*

Design science framework allows for rigorous application of IT tools to support the organizational strategy and infrastructure in the specific domain. This of course, requires following a set of activities or process to produce the artifact. Hevner (2004) provide guidelines for design science research as shown in Table 1. In what follows, we briefly discuss some of these outlines as they apply to our study:

The problem relevance and research contributions have been captured in the Sections 1 and 2 of this article respectively. We are proposing a smartphone-based monitoring tool (**Design as an Artifact**) to detect early signs of exercise-induced bronchoconstriction. Here we employ existing devices, algorithms and mathematical models as building blocks for the realization of the proposed system. The design model makes room for interaction of several IT techniques in mobile computing, which ultimately provides the required services by the monitoring tool as described in the system design (see Section 4).

Since the proposed system is going to be implemented in real-life situation, ***testing*** and ***observational*** methods are recommended for the design evaluation to ascertain its usability in the environment where the system is to be deployed.

In the testing method, we evaluate the functionality, completeness, consistency, and accuracy of the individual components of the system using existing tools – mobile phones with embedded sensors, computer systems, mathematical models and algorithms. The built-in sensors are primary source of real-time data collection. However, pre-recorded dataset obtained from a repository were used for the preliminary testing of the respiratory sounds analysis. The computer system provides the environment for statistical analysis in terms of sensitivity, specificity and accuracy of the algorithms. The testing process is performed in modules with respect to the various components of the system each linking the other as shown in Figure 1. An integration of the three components would be performed as a pilot test with volunteers.

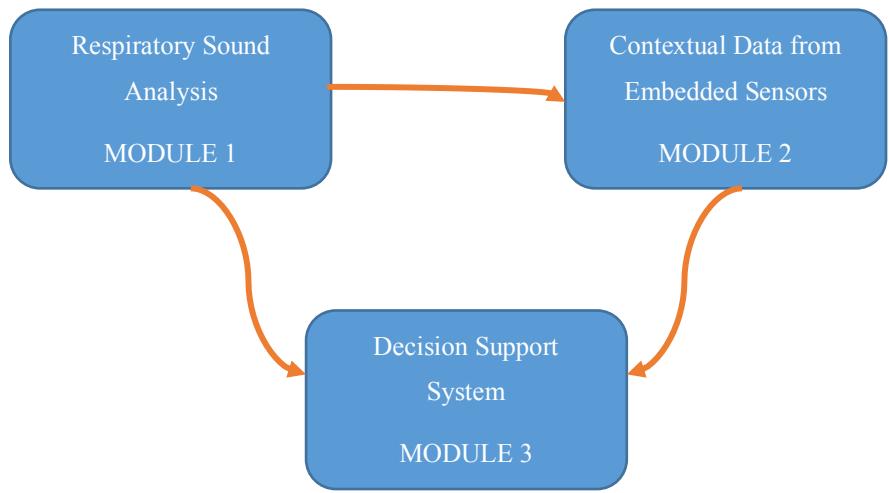

*Figure 1.*     *The three major components of the proposed monitoring system*

A case study approach is deemed appropriate in the observational method of the design evaluation, in order to provide the essential feedback – performance and reliability of the proposed system for its full implementation. The research focus group are the athletic youths and school-age children (10 -12 years) residing in rural areas in Jamaica. However the actual sample size is yet to be determined, given that the study is ongoing; and presently, we are consulting and collaborating with health experts in pulmonology and sports medicine to extract more knowledge for the system building. Data gathering tools we have employed so far in this regard include documentation review, interviews and observation.

Involvement of both design science and behavioral science approaches (vis à vis quantitative and qualitative methods) in the study provides basis for data triangulation which consequently, allows for more elaborate and deeper understanding of the phenomena being studied.

## 4. SYSTEM DESIGN

Figure 2 shows an overview of the proposed system architecture. The design comprises of both hardware and software components to realize an integrated monitoring system. The audio sensor, activity sensors and ambient sensors in the smartphone constitute the basic hardware for capturing both physiological and clinical data of the patient. The system architecture also includes data processing components which enable extraction and representation of important features of the data (Uwaoma & Mansingh, 2015). The DSS kernel houses data analysis components - classification algorithms and context recognition techniques, which provide analysis and description of context data sets, and also generate the required services for the user.

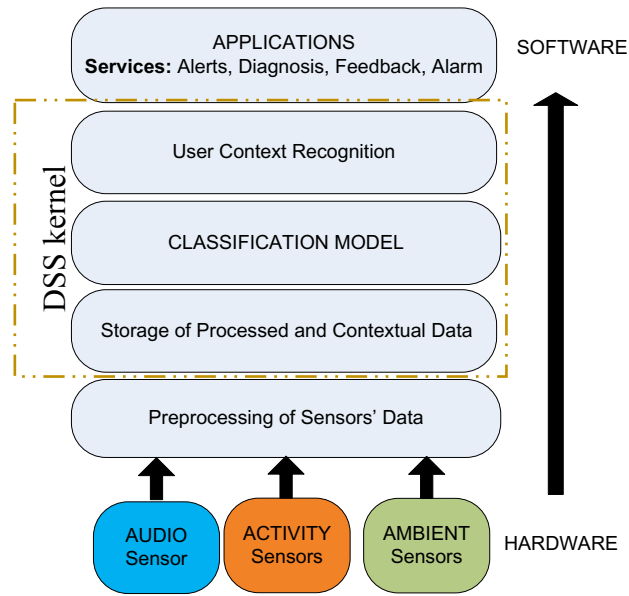

*Figure 2.    An Overview of the Monitoring System Architecture.*

The design is conceptualized based on the premises that:

- Mobile phones are capable of recording breath sounds and performing analysis on the recorded signal using computer algorithms.

- Wheeze detection and analysis provide medical doctors with critical information on how to adjust treatment of patients with asthma condition.

- Smartphones are able to correlate patterns in the user's movement and also recognize sudden changes in bodily position using embedded sensors like gyroscope, accelerometer and magnetometer. Table 2 provides brief description of available sensors in modern smartphones.

- Ambient data such as temperature, humidity and air pressure can be captured by built-in sensors to provide context on detected events.

- Expert systems can use the evaluation from the sensors' measurements to decide on the severity of a patients' condition.

| Sensor | Description |
|---|---|
| Gyroscope | Measures the orientation of a device in pitch, roll and yaw; and also the rotational rate of the device in x,y,z axes. |
| Accelerometer | Measures the acceleration force that is applied to the device including force of gravity. |
| Magnetometer | Measures the heading direction or the geomagnetic field of the device in three axes. |
| Barometer | Measures the ambient air pressure. |
| Hygrometer | Measures the humidity of ambient environment. |
| Thermometer | Measures the ambient (room) temperature. |
| MEMS Microphone | Records sound transmitted to the phone. |

*Table 2.    Embedded Sensors in Smartphones.*

## 5. DESIGN IMPLEMENTATION AND PRELIMINARY TESTING

The first phase of the design implementation requires two procedures for feature extraction from sensors' data namely; wheeze detection and activity recognition. The raw sensor data may need to be pre-processed on the firmware or at the application level before exposing them to the techniques used in the two subsystems.

**Wheeze Detection**

Wheezing is one of the frequent symptoms experienced by patients with asthma. Detecting the presence of wheeze, its duration and proportion in a breath cycle could assist physicians to ascertain the severity of attack. Wheeze detection systems are built on the basis of stethoscope principle. Several algorithms have been proposed for automated wheeze detection in respiratory sounds. These include Tonal Index [TI], Kurtosis [K], Energy Ratio [ER] (Wisniewski & Zielinski 2010), and Adaptive Respiratory Spectrum Correlation Coefficient [RSACC] (Yu et al. 2013).

 In our design, we implemented Time-Frequency Threshold Dependent (TFTD) algorithm which by extension is a modification of RSACC algorithm with low computational requirement for resource-constraint devices like smartphones. Characteristic features used for determining the presence of wheeze include the frequency harmonics, continuity and duration of wheeze intervals in a breath cycle. For the preliminary testing of the TFTD algorithm we used pre-recorded lung sounds obtained from a repository. An android phone was also used to record simulated 'wheezes' and normal breath from a healthy individual. The major operations in the algorithm include segmentation and decomposition of the breath signal to extract spectral and temporal features using Short Time Fourier Transform (STFT) in (1).

$$X[n] = \sum_{k=-\infty}^{\infty} s[n]w[n-m]e^{-j2\pi k/N} \tag{1}$$

Actual respiratory segments are isolated from the baseline signals using local minima and wheeze segments are identified by performing cross correlation (2) on adjacent respiratory segments. The proportion and duration of the detected wheezes is then determined from the respiratory cycle.

$$C(n) = \frac{\sum_{k=0}^{N}[(x(k)-mx) \ (y(k)-my)]}{(\sqrt{(x(k)-mx)^2}) \ (\sqrt{(y(k)-my)^2})} \tag{2}$$

The overall performance of the algorithm has a mean accuracy of 0.88±2 with false alarm rate of about 10.6%. The erroneous detection is attributed to strong correlation of respiratory signals between wheezing sounds and other respiratory sounds due to existence of peaks in other breath sounds which are highly similar to peaks representing wheezes. To further discriminate between wheezing sound and other respiratory sounds we included other quantitative and qualitative measures in the analysis. These include ranges of dominant frequencies and visualization of the sound waveform, spectrogram and cross correlation plot resulting from the TFTD algorithm (Uwaoma & Mansingh, 2015).

**Activity Recognition**

In this section we describe two quantities of interest in the activity recognition subsystem using the motion sensors. These features are patient's *posture change* obtained from the Orientation measurement and *activity level* provided by the linear motion parameter. With the device placed strategically on the body trunk, the linear movement is represented by the readings of the three accelerometer axes as follows:

z-axis: captures forward movement;

y-axis: captures upward/downward movement and x-axis: captures horizontal movement.

For posture change, the orientation or attitude data is measure in degrees (between -180 and 180). Evaluating the posture of the trunk after a patient has undergone a burst of energetic or vigorous exercise helps to ascertain if there are remarkable degree of posture variations in terms of a patient bending forward or tilting sideways in order to get sufficient air in the event of airway obstruction.

The Movement Intensity (MI) metric also known as average resultant acceleration measures the instantaneous intensity of patient's movements and it is generated from accelerometer readings of the three axes as show in (4).

$$\text{Movement Intensity (MI)} = \sqrt{(x_i^2 + y_i^2 + z_i^2)} \quad (4)$$

To determine the level of activity, linear accelerometer readings resulting from patient's movement are categorized into two:

(i) Stationary (e.g. sitting, standing, lying)   (ii) Ambulatory (e.g. walking, running, jogging).

To distinguish between the two categories of activities, we compute the movement intensity using average motion (AM) expressed as:

$$\text{Average Motion (AM)} = \frac{1}{N}\sum_{i=1}^{N} |MI_i - SMA_i| \quad (5)$$

Where N is the length of sampling/sliding window and SMA is the combined Signal Magnitude Area of the three axes computed as:

$$\text{Signal Magnitude Area (SMA)} = \frac{1}{N}\left(\sum_{i=1}^{N} |x[i]| + \sum_{i=1}^{N} |y[i]| + \sum_{i=1}^{N} |z[i]|\right) \quad (6)$$

SMA is a metric used to track or predict energy expenditure in everyday activity. It is also used to discriminate between a resting state and high intensity movements in a classification framework. Thus, using a threshold-based classification in a scenario where the patient is apparently NOT moving, the AM value is relatively low and where the activity status is "not resting" (noticeable strenuous movements), the AM value is relatively high.

In the following subsections, we describe strategies adopted to enhance the results obtained from the early implementation of the design. First we consider machine learning approach for classification of extracted features in the wheeze detection and activity recognition processes. We also consider using embedded intelligence to provide alerts to patients and their caregivers on deviation from baseline measurements. This will be accomplished by developing Expert Systems to integrate the rules and contextual information in order generate the required services to the user.

**Machine Learning Techniques**

Machine learning (ML) has become an integral part of detection and context recognition systems particularly in healthcare. ML techniques provide light-weight algorithms for event detection and context recognition to reduce energy consumption by the monitoring system. Furthermore, in the wheeze detection that was tested on recorded breath sounds, we noted "false positives" detection in breath signals that contains no wheeze. Advance classification models and machine learning techniques could be used to minimize the occurrences of erroneous detections. This of course will involve searching for optimal classification parameters from extracted signal features. There are several classification models used for optimal parameter search in extracted features. These include support vector machine (SVM), MPL ANN algorithm and Decision Trees. The choice of the model however, will depend on the accuracy, efficiency and low computational and power requirements on a mobile phone. Figure 3 depicts the process flow of the machine learning process.

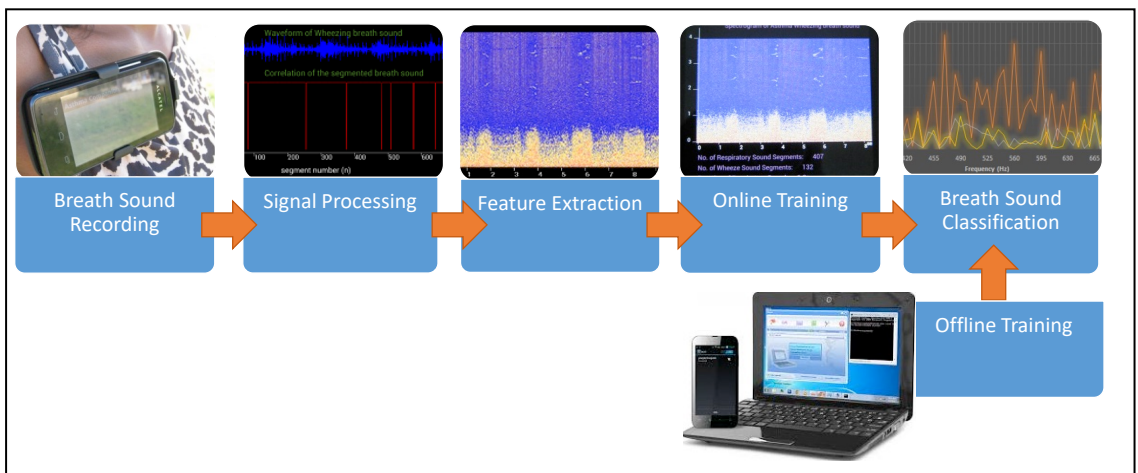

*Figure 3.    Process flow of the Machine Learning Technique*

**Embedded Expert Systems for Decision Making**

Expert systems (ES) designs are based on artificial intelligence techniques involving expert knowledge expressed through rules. Rule-based DSS for m-Health systems require context information for ad-hoc reasoning in order to reduce false positive alarms on a detected event. For exercise-induced asthma monitoring, contextual data such as patient's activity type (walking, running, jogging) and weather condition (ambient temperature, humidity) could help in quicker and correct recognition of asthma flares given the presence of potential triggers. Information on patient's body postures (sitting, lying, standing, tilting forward, and tilting sideways) could also improve the accuracy in determining the severity of the attack.

A typical rule-based DSS for healthcare comprises of three main components namely:

- Fact base which contains physiological and contextual data from patients
- Inference Engine which handles the decision process
- Rule base which contains expert knowledge that determines what action or service is to be executed/provided.

These three components are represented as layers in a simplified DSS model as shown in Figure 4.

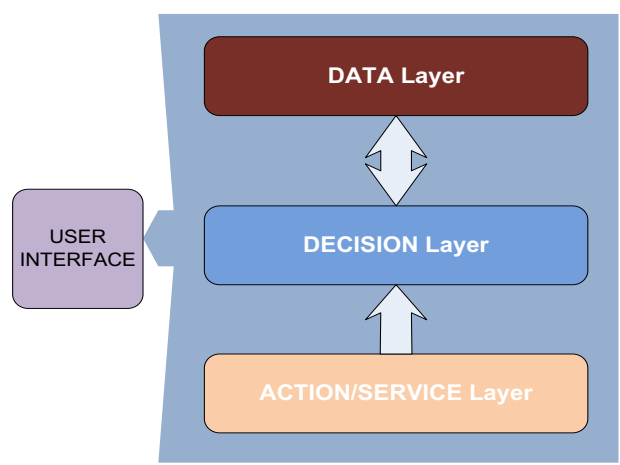

*Figure 4.     The proposed Decision Support System Model*

In our design we intend to combine production rules with ontologies in the DSS model for efficient processing of information specific to the problem domain. Production rules play important role in expert systems as they provide the basis for the computational model to draw deductions from the presented scenario. They are often structured using the IF-THEN statement. Ontology on the other hand, defines structures and descriptive logics used to establish and actively reason upon relationships among the structures in the domain. New relationships could also be discovered while querying the existing ones in a given ontology. Figure 5 shows a sample of ontological structure design for the specific issue we are considering which is Exercise-Induced Asthma (EIA).

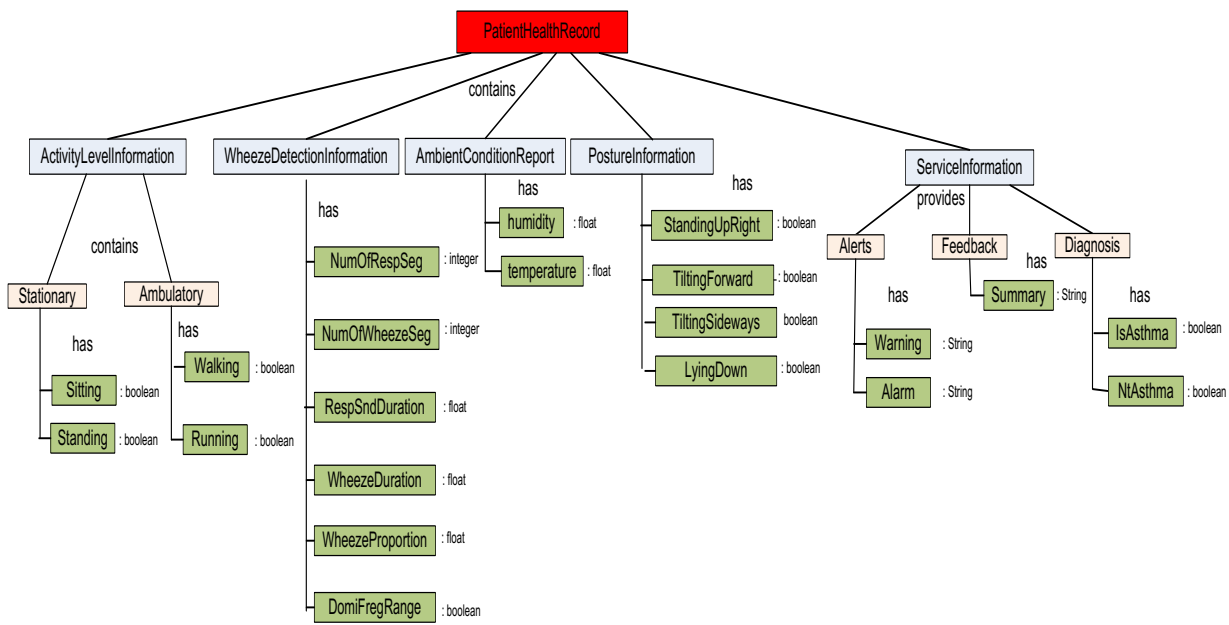

*Figure 5.      Ontological Structure of the proposed DSS*

## 6.   APPLICATION OF THE PROPOSED SYSTEM IN A RURAL SETTING

There is no doubt that ownership of mobile phones these days has become a household necessity rather than a luxury as the "usage captures the entire age spectrum from school children to the elderly" in both developed and developing countries (Boulos et al., 2011). It is even more obvious that the importance of these communication devices transcends long distance conversation as they are now being deployed in various areas of human endeavor to solve peculiar problems relating to accounting, surveillance, health, security, etc. The significant drop in the prices of smartphones also makes it more affordable and available to people with low-income (Boulos et al., 2011).

Here, we describe a case study of the application of the asthma monitoring system in a remote area with limited access to medical facilities. As we mentioned earlier, the proposed asthma monitoring system is solely based on smartphone platform and given the ubiquity of these intelligent devices, the monitoring tool as standalone device could be useful in remote and rural areas where internet connection and mobile data services may not always be present. An instance of the system application can be situated in a rural school setting during PE classes or sporting activities. Assume an asthmatic pupil forgot to take his asthma preventive medications prior to the exercise but has the monitoring device on him; the system could detect and alert the pupil and his caregiver (teacher) on unfavorable weather condition or activity level that could predispose an asthma attack. Even if the pupil ignores early warnings of asthma flare, it behooves the teacher or caregiver to recall the pupil from field or send him for immediate medical attention to avert possible escalation of the symptoms or hospitalization.

## 7. CONCLUSION

The overall goal of the system design is to develop real-time application for monitoring and detection of early symptoms of asthma during exercise or rigorous activity. The envisaged system would alert users on detection of any anomaly so as to avoid further exacerbation of the condition; and also generate feedback on patient's health status. We have implemented the algorithms for breath sound analysis and we are currently working on activity recognition using advanced classification algorithms performed on a smartphone. As m-Health is gaining wider acceptance in modern medicine, inclusion of domain medical knowledge in decision making implemented in expert systems as part of the design, will help provide right diagnosis and treatment for the patient by health professionals. Thus, having all the operations performed exclusively on the mobile device, we believe our design approach will not only improve on the usability and reliability of the previous systems; but also serve as a convenient tool for quick medical interventions in remote areas where specialized clinical devices and services are not easily accessible.